\documentclass[preprint,12pt,authoryear]{elsarticle}

\usepackage{graphicx}
\usepackage{amssymb}
\usepackage{comment}

\newcommand{\proposedmethod}{Panacea~}{}
\newcommand{\proposedmethodz}{Panacea}{}
\newcommand{\proposedfullname}{Polarizing Attributes for Network Analysis of Correlation on Entities Association }{}

\newcommand{\proposedcompany}{Panasonic~}{}
{}

\journal{arXiv}

\begin{document}

\begin{frontmatter}



\title{
Visual Exploration System for Analyzing Trends in Annual Recruitment Using Time-varying Graphs
}

\author[add1]{Toshiyuki T. Yokoyama\corref{correspondingauthor}}

\address[add1]{The University of Tokyo, Japan}

\author[add2]{Masashi Okada}
\author[add3,add2]{Tadahiro Taniguchi}
\address[add2]{Panasonic Corporation, Japan}
\address[add3]{Ritsumeikan University, Japan}

\begin{abstract}
Annual recruitment data of new graduates are manually analyzed by human resources specialists (HR) in industries, which signifies the need to evaluate the recruitment strategy of HR specialists. Every year, different applicants send in job applications to companies. The relationships between applicants' attributes (e.g., English skill or academic credential) can be used to analyze the changes in recruitment trends across multiple years' data. However, most attributes are unnormalized and thus require thorough preprocessing. Such unnormalized data hinder the effective comparison of the relationship between applicants in the early stage of data analysis. Thus, a visual exploration system is highly needed to gain insight from the overview of the relationship between applicants across multiple years.
In this study, we propose the \proposedfullname (\proposedmethodz) visualization system. The proposed system integrates a time-varying graph model and dynamic graph visualization for heterogeneous tabular data.
Using this system, human resource specialists can interactively inspect the relationships between two attributes of prospective employees across multiple years. Further, we demonstrate the usability of \proposedmethod with representative examples for finding hidden trends in real-world datasets and then describe HR specialists' feedback obtained throughout \proposedmethodz's development. The proposed \proposedmethod system enables HR specialists to visually explore the annual recruitment of new graduates.

\end{abstract}



\begin{keyword}
Human resources \sep Data visualization \sep Property graph \sep Time series analysis


\end{keyword}

\end{frontmatter}

\section{Introduction}
Recruitment of new employees is one of the most vital duties in Human Resources (HR) management. HR specialists themselves wish to discover the comparative and chronological trends of an applicant from the pool of applicants' historical data. For example, they wish to compare distributions in the English skills of prospective employees. However, the heterogeneity of the large database requires a great deal of pre-processing before the trend analysis, resulting in actual data loss. A method to gain insight into the relationships over multiple years of database records required. An interactive visualization platform is therefore essential for the interactive exploration of data by HR specialists.

Data analysis conducted by a company facilitates the evaluation of previous business strategies and the discovery of hidden trends or biases \citep{chen_business_2012_2}. The importance of data analysis is also recognized in HR management \citep{chien_data_2008, xiaofan_application_2010}. Among various HR functions, recruitment is one of the most important tasks for growing a company and assigning appropriate personnel to each section, department, etc. For most local companies in Japan, the standard recruitment source is a short annual recruitment period for new graduates each year \citep{pucik_white-collar_1984, peltokorpi_recruitment_2016}\footnote{Around the same time, most Japanese local companies recruit new graduates in a short period every year, usually for a few months. At the same time, students in their final year of school apply for one or more companies during this period. Companies select prospective employees among the applicants through a screening process, including a series of interviews. As a result, large companies have to process many applications in a short period.}.
It has been estimated that more than half of university graduates have worked in the same company for more than a decade, as a result of the lifetime employment system in large Japanese companies \citep{ono_lifetime_2010}. Therefore, an effective review of prospective employees into a company is a critical task for HR specialists.

Analyzing a large volume and a wide variety of applicants' information is challenging. Applicants' data are stored in the recruitment database, as records in a table. Different data types are used to store attributes of applicants' data in each column of the table in the database (e.g. name as a string, English exam score as a number, or academic credentials as a category, etc.). Sometimes attributes fields are left empty and not normalized across the table, and attributes have different data types. 

We denote this characteristic of unnormalized attributes as \textit{heterogeneity}, which makes it difficult to compare the attributes of applicants across different years. Usually, larger companies receive more than a thousand applications each year, which increases pre-processing effort. A large number of applicant records are managed in an recruitment management system, but rich analytical functions are excluded from the system's design. Therefore, HR specialists must manually review the applicants' data.

Resolving the heterogeneity of attributes involves a lot of pre-processing effort, which is still a challenging part of the analysis workflow \citep{milani_visualization_2020}. Without pre-processing, spreadsheet and business intelligence (BI) tools do not provide efficient aggregation or visualization. HR specialists wish to extract several attributes from applicants' data to focus on further trend analysis rather than to spend much time normalizing the data. A visualization system is the most effective way to address this issue without writing any code. 

With the discussion with several HR specialists, their requirements for the visualization system are as follows, (A) A user should be able to choose which attributes to use for further analysis, (B) A user does not want to miss out a smaller number of attributes, i.e., rare cases which are often excluded in quantitative analyses, (C) A user can compare attributes across years, and (D) A user can explore an overview of the data based on user's criteria. Without these four requirements, for example, finding multi-year trends on prospective employees applying for a position in the HR department becomes difficult, since they represent a minute proportion of the entire prospective employees. To the best of our knowledge, there is no existing tool that satisfies these HR data analysis requirements.

\begin{figure*}[tb]
 \centering 
 \includegraphics[width=\textwidth]{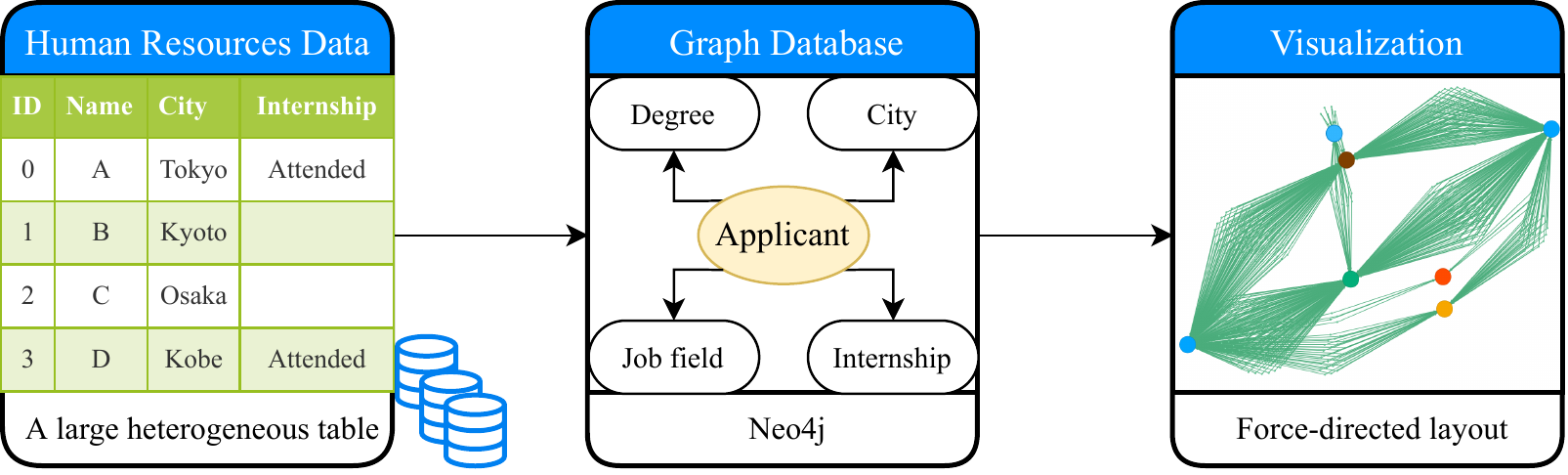}
 \caption{Data processing pipeline of the system. Input is a large data table of a recruitment database. We convert the input data as a single property graph stored in a graph database. The graph describes the relationship between prospective employees and their attributes such as skill, age, and university. The proposed method is a web application that visualizes the relationship between two attributes of applicants from the graph database.}
 \label{fig:interface}
\end{figure*}

Here, we propose the \proposedfullname (\proposedmethodz) visualization system. The system provides an interactive interface to explore every combination of two attributes of prospective employees each year. We employ Property Graph (PG) as the data structure to provide more intuitive visualization for heterogeneous data. The proposed system satisfies the requirements (A)-(D), which we describe in the design requirement section. 
The workflow for running the system is as follows (\autoref{fig:interface}): (1) convert the tabular data to a time-varying graph data; (2) extract subgraphs by user-specified attributes and years; (3) draw the subgraph on an interactive interface. 

To achieve this, a combination of visualization of dynamic time-varying graph and data-wrangling method to convert tabular data into a time-varying graph is needed. The idea to convert tabular data into a graph has already being explored \citep{heer_orion_2014, liu_ploceus_2014,8019835,8986909,SHI202043}, but however, existing data wrangling methods do not explicitly encode timepoints on a graph. Meanwhile, dynamic graph visualization tools are not responsible for a graph conversion from temporal tabular data. We define a multi-partite graph to represent time-varying information of tabular data and then design an interactive interface to visualize dynamic graphs. Our primary contributions are:

\begin{itemize}
    \item Time-varying graph modeling and dynamic visualization system for heterogeneous tabular data with time scale.
    \item An interactive interface with dynamic graph visualization that satisfies the requirements on annual recruitment analysis.
    \item Three case studies and user studies demonstrating how HR specialists use the proposed system to analyze the database by finding unique trends.
\end{itemize}

\section{Related Work}\label{sec:related}
\paragraph{Visualization for Human Resources} 

The HR data are stored in a tabular form in a relational database (RDB). To inspect such data, spreadsheet tools, such as Microsoft Excel and Google Spreadsheet, would be the first choice. Those tools are often used to display the original data and perform fundamental aggregation, such as sorting, filtering, and visualizing using predefined charts. For example, Microsoft Excel is the most frequently used tool to manipulate tabular data for HR analysis \citep{lunsford_tools_2018}. Spreadsheet tools can compare two attributes at a point of time or they can show the time-series variation of one type of attribute. However, such tools with the two-dimensional table representation are unsuitable for analyzing two attributes with time scale simultaneously. Moreover, spreadsheet tools are not suitable for such heterogeneous or large tabular data due to aggregation and performance issues; thus, every time users must extract tables for their purposes.

Recently, BI tools such as Tableau, SpagoBI, and Qilk have been used for business analytics \citep{gounder_survey_2016, morton_dynamic_2012}. For HR analysis, \cite{kale_big_2016} used Tableau to analyze trends in job descriptions in New York with bar and line charts. While BI tools are useful for aggregation, there are two concerns with BI tools. First, the backends of BI tools are based on the RDB model, thus the data must be normalized. We often find that tremendous effort has been spent to normalize the data.
Second, once users specify the attributes to be analyzed, the tool can visualize the attributes on a sophisticated interface. Users will only use BI tools effectively if they can normalize all data and know what kinds of visualization and aggregation will be beneficial. Before starting tabular analysis and visualization, users need to decide what columns to focus on.

If users have python skills, they can use Jupyter Notebook \citep{Kluyver:2016aa} for inspecting data. Various kinds of visualization and inspection libraries are also available in python. However, those libraries require more programming skill, thus hindering non-programmers from inspecting data by themselves.

Data heterogeneity is a major challenge for information visualization systems \citep{liu_survey_2014}, and that of the HR data analysis system is no exception. How to visualize HR data has been being explored, and several visualization systems for HR data have been developed. Proactive \citep{lee_fighting_2007} is a job description search engine with a rich interface. It displays a list of job descriptions in a tabular manner; however, the relevance between records is not shown on the interface. \cite{walter_implementation_2017} presented a system that displays each applicant with various metrics, such as talent or performance. The system is useful in allowing a user to know the attributes of each applicant; however, the relationships among attributes are not shown. In contrast, the proposed system focuses on the relationships among attributes of prospective employees.

\paragraph{Visual Analysis on Temporal Data}

If we wish to visualize the time flow of a single type of attribute, a two-dimensional flow-based temporal data visualization, e.g., bar chart or line chart, is a widely accepted method \citep{krueger_traveldiff:_2016, lu_visual_2019}. 
In flow-based visualization, the X-axis is the time, and the Y-axis is the one-dimensional scalar value of the attribute, generally. The flow-based approach has a variant with additional information on the Y-axis. For example, the tracking graphs method visualizes each timepoint of nodes that vertically stacked up against the X-axis \citep{widanagamaachchi_interactive_2012}. We utilize a flow-based visualization technique to visualize an overview of a given attribute and timepoints. However, two-dimensional flow-based visualization has two disadvantages. First, two-dimensional flow-based visualization does not visualize both temporal relationships and the relationships between attributes for each timepoint. Second, flow-based visualization requires the linearization of the position of attributes when attributes are encoded as a scalar value on the Y-axis. We will also review the way to visualize both temporal and two-dimensional information in the following subsections.

\paragraph{Visual Analysis on Categorical Data}
We see that many categorical, albeit unnormalized, data are stored in the HR database. State-of-the-art techniques of categorical data visualization were reviewed by \cite{alsallakh_state---art_2016}. They categorized these techniques as Euler-based, overlays, node-link, matrix, aggregation, and scatter. The node-link diagram, also known as network graph, is used to depict categories and their elements as nodes and edges in a graph. The node-link diagram can visualize either the relationship between categories and elements as bipartite graph \citep{misue_anchored_2007, dork_pivotpaths_2012,alsallakh_radial_2013}, or between categories as Parallel Sets \citep{kosara_parallel_2006}.

They described the advantage of the node-link diagram as highlighting elements as nodes and clustering nodes of each category, making it easy to understand. Radial Sets \citep{alsallakh_radial_2013} locate the category nodes along the arc of the circle. Element nodes are located inside the circle, which belong to multiple categories.
Parallel Sets \citep{kosara_parallel_2006} visualize categories as parallel coordinates plot and the frequencies of the combination of categories as edges between categories with an interactive interface. They aim to visualize complex data information in its entirety. However, it may be difficult to grasp the description for those who see it for the first time. 
The main downside of a node-link diagram is that crossing edges make it difficult to understand the diagram. Also, the number of edges in the graph is often limited to about hundreds because of increasing clutter \citep{alsallakh_state---art_2016}.
Nevertheless, the node-link diagram can be integrated with dynamic graph visualization for temporal data. We utilize the node-link diagram for visualization of categorical heterogeneous tabular data.

\paragraph{Visualization for Dynamic Graphs}

Several static graph visualization tools for general purpose, including Cytoscape, GraphViz, and Gephi \citep{shannon_cytoscape:_2003, farin_graphviz_2004, bastian_gephi_2009}, provide a sophisticated way to visualize any type of graphs; however, these tools were not designed to render dynamic graphs. The review articles by \cite{beck_state_2014,beck_taxonomy_2017} described a hierarchical taxonomy for categorizing visualization techniques for dynamic graphs. It can be subdivided into four types: (1) timeline node-link, (2) timeline matrix, (3) animation with a special-purpose layout, and (4) animation with a general-purpose layout. Hereafter, we describe each layout. 
(1) Timeline node-link approaches, such as the work of \cite{greilich_visualizing_2009} and \cite{burch_parallel_2011}, 
show superimposed or juxtaposed nodes for each year, which visualizes the relationship between nodes rather than the graph topology. Small multiples are categorized as the timeline node-link approach. 
(2) Timeline matrix approaches, such as the work of \cite{burch_matrix-based_2013} and \cite{stein_pixel-oriented_2010}, 
are suitable for dense graphs; however, our target graphs are not dense as can be seen by the construction method. Timeline approaches visualize the time scale on the view, which requires a summarization of each time step. (3) Animation with a special-purpose layout can visualize graphs of each time step; however, the approaches need an abstract representation of nodes based on hierarchy or clusters. (4) Animation with a general-purpose layout can visualize graphs by general methods, and users can easily trace the position of nodes in each time step \citep{huang_-line_1998,hayashi_initial_2013}. We utilize (4) the animation general-purpose layout in the proposed system because it provides the most flexible visualization without any abstraction for each time step.

For selecting the method for visualizing dynamic graphs, how each method preserves the user's mental image of the graph, i.e., mental map, is an important criterion \citep{beck_state_2014}.
The animation approach is one of the effective methods that provide a mental map by maintaining coherency among time steps \citep{archambault_map_2013,archambault_can_2016}. The animation approach enables users to gain more insights from changes between subsequent years in several cases, as demonstrated by \cite{boyandin_qualitative_2012}. However, as described by \cite{hajij_visual_2018}, the limitation is that the graph topology can be lost on each time point. Several approaches to mitigate this limitation are available. For example, GraphDiaries \citep{bach_graphdiaries:_2014} displays an animated transition of graphs between time steps, where disappearing nodes are highlighted first, and then appearing nodes are highlighted. Similarly, TempoVis \citep{ahn_temporal_2011} displays the color difference between appearing and disappearing nodes.
Using a force-directed approach also mitigates the overhead of transitions identification through time steps \citep{kumar_visual_2006}. This is because this approach is able to trace the moving positions of the nodes of the previous timepoint. We utilize the force-directed approach for tracking the position of nodes.

\paragraph{Graph Visualization for Tabular Data}

The efficiency of graph-based visualization for tabular data has been discussed in Orion \citep{heer_orion_2014}, Ploceus \citep{liu_ploceus_2014}, Graphiti \citep{8019835}, Origraph \citep{8986909}, and Oniongraph \citep{SHI202043}. These are data-wrangling tools to convert tabular data into graph and provide a graph visualization interface. These tools based on the idea that the network structure in tabular data can be represented as graphs, and they support constructing a graph from tabular data interactively. Orion uses domain-specific languages and visual interface to display a node-link diagram without any attributes. Ploceus and following tools explicitly support to attach attributes on each node. Further, several tools also support a multi-layer graph. For example, Graphiti handles a multi-layer graph whose layers have different types of edges. OnionGraph uses node aggregation for hierarchical abstraction and provides filtering function for nodes or edges. Both tools assume that the topology among layers is maintained. However, time-varying graph does not assure that the graph topology is maintained through multiple time points. Therefore, special care to maintain a mental map is needed.

The temporal graph visualization has been adopted for some special cases. For example, MatrixFlow \citep{aec92e3bcea44bca96e923f35a43e235} is used for medical data and visualizes temporal network as timeline matrix approach. ecoxight \citep{10.1145/3185047} visualizes business ecosystems as an animated timeline node-link diagram, though the input must be a graph.

\proposedmethod shares the same idea with these tools to handle tabular data as a graph, but however, we especially focus on the temporal data visualization for temporal tabular data. In this study, we wrote custom scripts to convert the graph data and developed a visualization system to explicitly display temporal information on the node-link diagram.

\paragraph{Graph Database}

Which graph database we use is a design decision. Categorical data visualization by a node-link diagram internally converts categorical data into a graph-based representation. Storing all categories into a graph database is a persistent and scalable way to query from the frontend every time. To store graph data in a graph-based database, mainly two definitions of graphs are available, i.e., the resource description framework (RDF) \citep{brickley2004rdf} and property graph (PG) \citep{matsumoto_mapping_2018}, to store them on the database. RDF is a normalized data structure to describe relations between entities. PG is a more flexible graph format and is compatible with several graph databases, including Neo4j\footnote{https://neo4j.com}. RDF is well-normalized but occasionally complex to visualize and to query because all entities must be nodes. Therefore, in the proposed system, we utilize PG as the data format to store graphs.

Further, we employed Neo4j as a backend graph database. Neo4j has a company-provided visualization tool, i.e., the Neo4j browser. However, we needed to hold the position of specified nodes for preserving the mental map. Since the Neo4j browser does not support this feature, we wrote a custom frontend with vis.js\footnote{https://visjs.org}. Several studies have implemented their own visualization system on top of Neo4j. For example, \cite{onoue_development_2018} employed Neo4j as a backend graph database, and \cite{caldarola_improving_2016} used Cytoscape but stored data using Neo4j. Neo4j is state-of-the-art software to employ as the backend of the visualization system. We thus implemented our custom visualization modules on top of Neo4j.

\begin{figure*}[tb]
 \centering % avoid the use of \begin{center}...\end{center} and use \centering instead (more compact)
 \includegraphics[width=\textwidth]{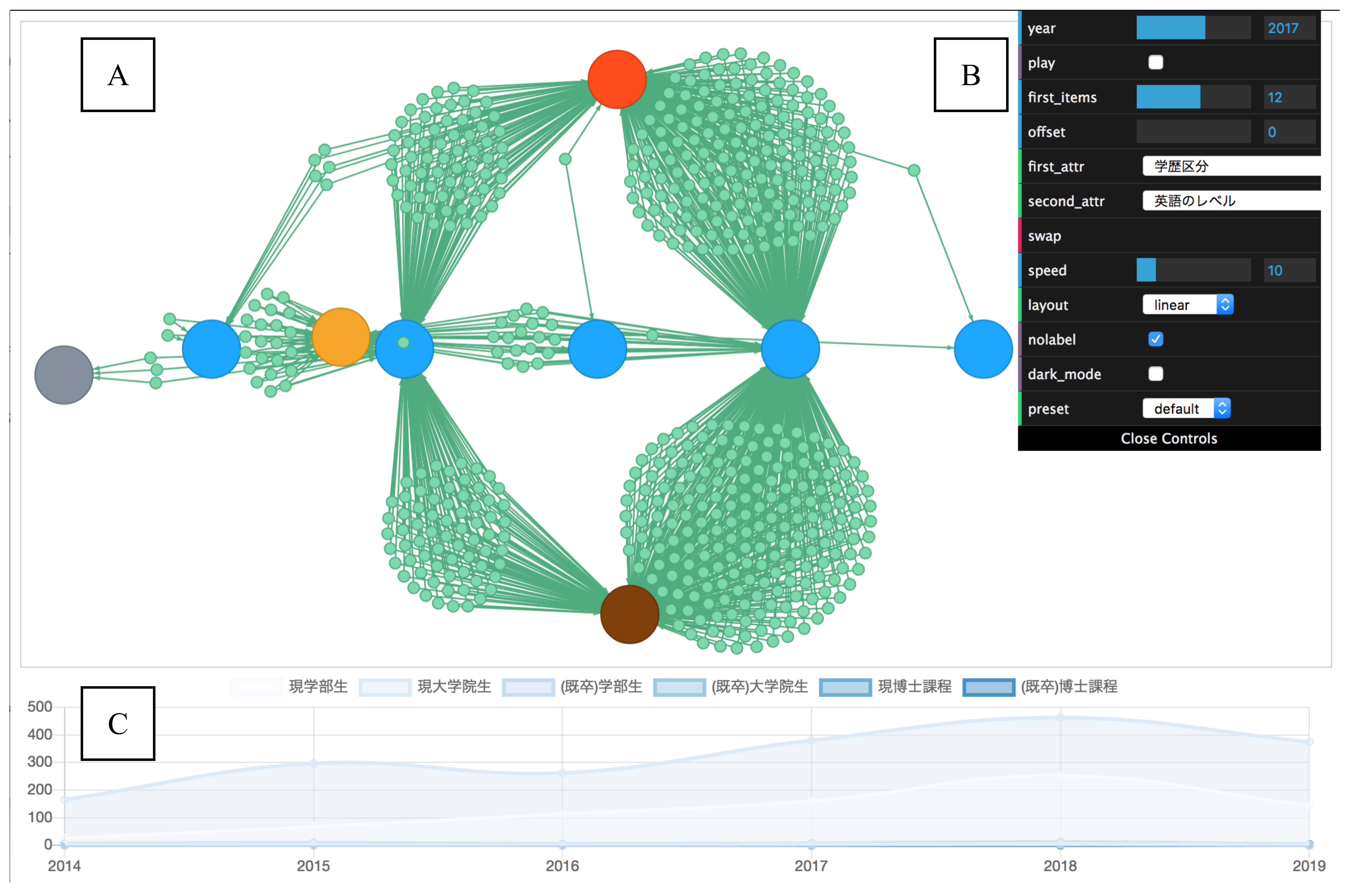}
 \caption{Screenshot of \proposedmethodz. The interface has graph (A), configuration (B), and chart (C) views. The graph view shows an example of the relationship between two attributes. The small green nodes are applicants, large blue nodes are primary attributes, and other large nodes are secondary attributes. Green edges are relationships between applicants and attributes.}
 \label{fig:screenshot}
\end{figure*}

\section{Design Requirements}\label{sec:design}

The primary goal of annual recruitment data analytics is to find trends or biases in prospective employees' historical data over different years. We invited three HR specialists from \proposedcompany Corporation to biweekly discussions (1-1.5 hours per meeting) on user requirement collection and prototype evaluations. We updated the system iteratively, which facilitated quick access to feedback. Since different HR specialists have different goals for analyzing data, there is no typical analysis workflow. From the series of interviews with HR specialists, we extracted common procedures and summarized them as follows: (1) extracting subtables by specified attributes, years, and/or applicants; (2) data analysis and visualization of the extracted subtables; (3) discussion on the results. Among that, (1) is a major obstacle for HR specialists because the heterogeneity of table columns requires much pre-processing effort and makes it difficult to have an overview. HR specialists demand a system that helps them extract subtables without writing any code.

Further, we identified four challenges of extracting subtables. (A) It is difficult to decide on which columns to focus on when extracting subtables. The combination of two columns is much larger than the number of the column;
thus, it is difficult for even HR specialists to know which columns to use in subsequent analysis. 
HR specialists require an interactive system to examine the entire table prior to extracting subtables. (B) Pre-processing might cause abstraction or summarization of data, thus obscuring the relationship between the original data and the visualization. Such methods often ignore the smaller number of attributes. However, these ignored attributes, i.e., rare cases or outliers, are sometimes important for HR analysis. For example, the number of applicants who can speak multiple languages is not high; however, such skills are valuable when a company seeks to expand its business to global markets. This prevents further exploration into the original data when HR specialists wish to gain more insight.
HR specialists require a visualization system that preserves the original data to avoid omitting the smaller-number attributes. (C) Applicants differ between years; however, most attributes are similar between years. HR specialists wish to analyze changes in the distribution of attributes over the years to identify trends or biases. (D) Different HR specialists have different criteria to perform analysis, thus a criterion that regulates the visualization must be customizable. For example, the attributes clustering pattern depends on the choice of each HR expert. As a summary of (A)-(D), HR specialists require an integrated system with (A) a bird's eye view that (B) does not omit outliers and (C) a time-varying view permitting (D) user customization. 

Herein, we discuss a visualization method that we employed in the proposed system. For (A) a bird's eye view (B) without omitting outliers, the advantage on a node-link diagram, which can highlight elements and which is easy to understand, is indispensable. Using the node-link diagram, nodes can be visualized without aggregation. Spreadsheet or BI tools partly support these requirements but require pre-processing. Matrix, scatter plot, and flow-based visualizations provide a one-versus-one comparison, e.g., one attribute versus another attribute, or one attribute on the time axis. In that case, the target attributes must be mapped to a single axis to align them horizontally or vertically, thereby incurring actual data loss in the relationship between attributes. These data should be handled with minimal modification or aggregation from the data stored in the database. Therefore, we employ PG as a data structure and a node-link diagram for categorical data visualization. For providing (C) a time-varying view , there are several options to integrate into the graph visualization, e.g. timeline (small multiples) or animation. Among them, the animation approach enables users to trace the transition between even distant years, thus keeping their mental map and providing them with more findings \citep{boyandin_qualitative_2012, archambault_map_2013}. In addition, we utilize the flow-based chart for an overview of the time scale. At last, enabling users to move the positions of the nodes corresponds to (D) user customization.
The entire system is described in the next section.

\begin{figure*}[tb]
 \centering % avoid the use of \begin{center}...\end{center} and use \centering instead (more compact)
 \includegraphics[width=\textwidth]{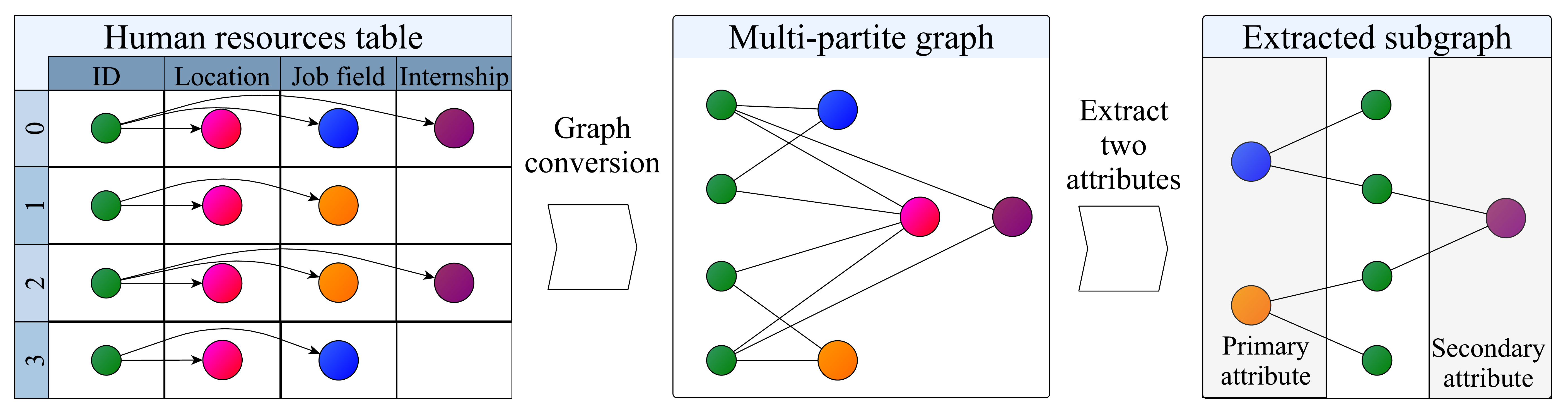}
 \caption{Data model in \proposedmethodz. First, the pre-processing script converts all tabular data into a single multi-partite graph. Next, users select two attributes and then query the backend server. The backend server returns a subgraph, including records of applicants and two corresponding attributes from the backend server.}
 \label{fig:datamodel}
\end{figure*}

\section{Implementation}

The proposed \proposedmethod system is a web application with three components (\autoref{fig:interface}): data pre-processing, a backend server, and an interactive frontend. For data pre-processing, we propose a data model based on a multi-partite graph and write custom scripts to convert a table to graph representation (\autoref{fig:datamodel}). Here, we use PG exchange format as an intermediate output of data processing \citep{chiba_property_2019}. Further, we employ Neo4j as a backend server and X2\footnote{https://github.com/g2glab/x2} as middleware for visualization. JavaScript and vis.js are used as the visualization library on the frontend.

\subsection{Data Model and Preprocessing}\label{processing}

Multi-partite graph G is expressed as $G = ((A, V_1, V_2, \ldots, V_n), E)$, where $E \subseteq A \times V_i$ such that $A$ represents applicants, each of $V_1, V_2, \ldots, V_n$ is an applicant attribute, and $E$ represents the relationships between an attribute and applicant. Note that each record in the table corresponds to a single applicant node. There is an edge $e \in E$ if and only if the applicant $a \in A$ has an attribute $v \in V_n$. From the frontend, users select two attributes $V_x, V_y$ from $V_1, V_2, \ldots, V_n$. Queries to retrieve subgraphs are encoded in Cypher (a query language for Neo4j).
Based on the definition of PG, nodes can have properties. Here we set \textit{type} property of each attribute node to describe the category name of attributes, e.g. \textit{academic credential} or \textit{internship history}. We also set \textit{year} property of each applicant node to specify the year to explicitly encode time scale on a graph.

Since we have already determined the design requirements, defining a custom conversion to satisfy these requirements is more reasonable than writing domain-specific languages or using data wrangling tools. Indeed, manual curation using existing data wrangling tools is not practically suitable to convert large and temporal tabular data into a time-varying graph. Therefore, we implement novel custom scripts to encode the temporal tabular data into a time-varying multi-partite graph.

In the HR database, each row is an applicant and each column is an attribute. We simply regard each column as either (1) a single-column attribute, (2) a property of the applicant node, or (3) a multi-column attribute. For example, \textit{academic credential} should be assigned as (1). \textit{Name} should be (2) because \textit{name} is tightly linked to the attribute; therefore we do not want to regard \textit{name} as an independent attribute. Let us consider \textit{internship history} for example of (3). There are three columns named \textit{internship history1}, \textit{internship history2}, and \textit{internship history3}. These columns contain company names as string datatype where applicants worked as an intern. These columns should not assign as independent attributes because the three columns are just an inflated array of \textit{internship histories}. Thus, we assign these columns to the same type on attribute nodes to merge these columns into one category\footnote{Such columns represent poor database design because the columns are not normalized. Instead, we should have an internship company table with a unique key for each company and an intermediate table with two keys for applicants and companies. However, we could not modify the original table structure due to the limitations of the recruitment management system.}. Most of the columns are categorized as (1), but we find that several columns should be categorized as (2) or (3). There are several models to convert tabular data to PG \citep{10.1145/2484425.2484426, 10.1145/3035918.3035949}; however, those models do not convert (3) a multi-column attribute. The advantage of graph representation is that graph can handle such kinds of unnormalized relational data smoothly.

The entire procedure to convert the HR data into the graph is as follows. An empty graph $G$ is initialized, and the following procedure is repeated for each record in the table: First, the applicant node $a \in A$ is inserted into $G$ with properties from all elements that are a property (2). Next, for all elements that are an attribute node (1) or (3), a tuple of two nodes and an edge ($a$, $e$, $v$) $\mbox{ s.t. }$ $a \in A$, $e \in E$, $v \in V_1, \ldots, V_n$ is inserted into $G$. At last, graph $G$ is imported into Neo4j.

\subsection{Overview of Systems}

\proposedmethodz's frontend has three view modules, i.e., graph, configuration, and chart views, as shown in \autoref{fig:screenshot}.
The chart and graph views work complimentary. The chart view visualizes one axis and temporal information through the whole timepoint. The graph view visualizes two attributes and temporal information between arbitrary two timepoint. With the combination of components, we provide ``overview first, zoom and filter, then details-on-demand" system introduced in Shneiderman's Visual Information-Seeking Mantra \citep{shneiderman_eyes_1996}. We can visualize two attributes through whole timepoint. Two attributes are classified as primary and secondary attributes. The primary attribute is highlighted as their number or position on configuration and graph view, thus working as a criterion to be compared with the secondary attribute.

\paragraph{Graph View}
\begin{figure}[t!]
 \centering % avoid the use of \begin{center}...\end{center} and use \centering instead (more compact)
 \includegraphics[width=\columnwidth]{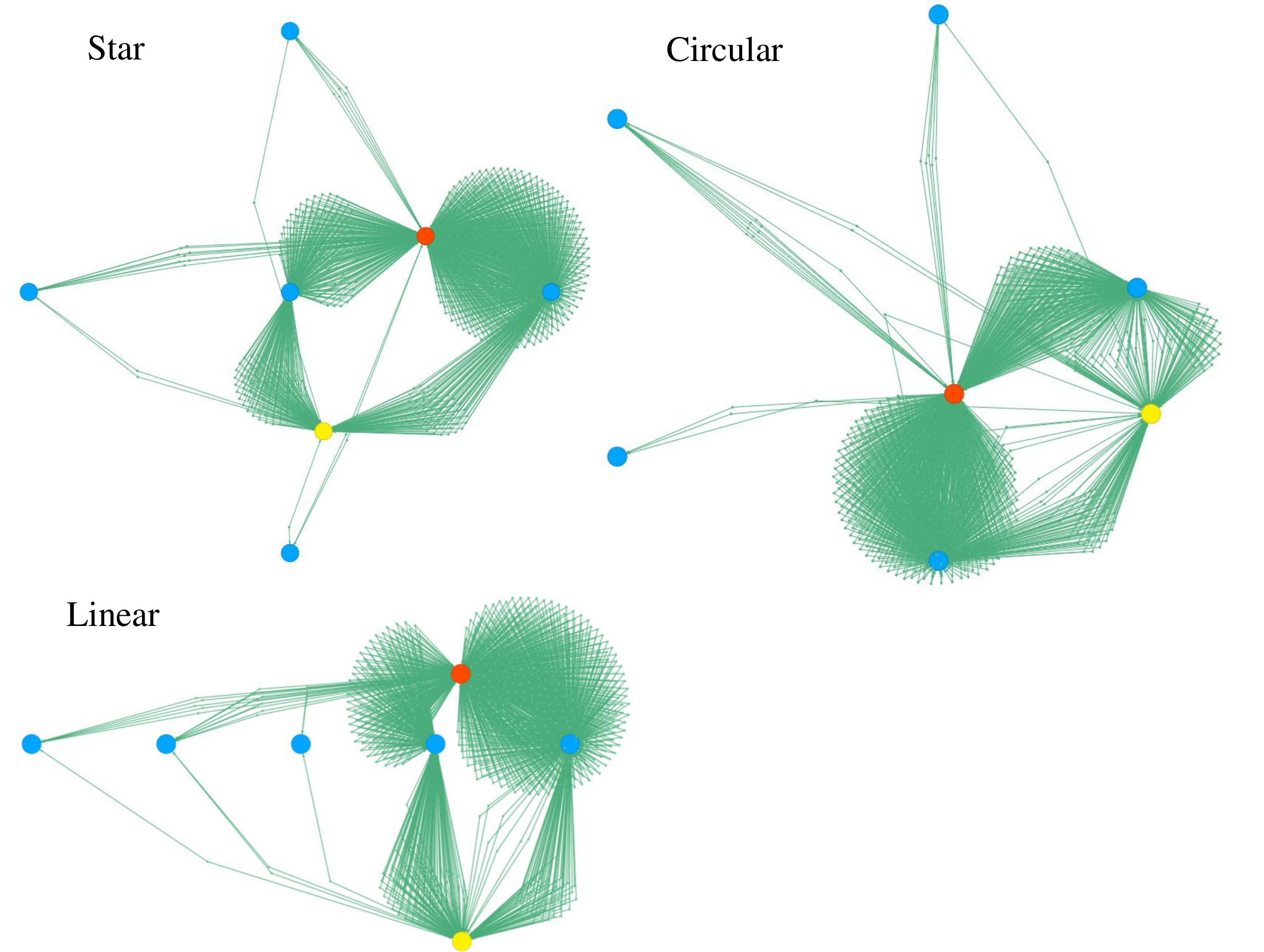}
 \caption{Three types of initial node layout implemented in \proposedmethodz.}
 \label{fig:layout}
\end{figure}

The subgraph $G(year, x, y) = ((A_{year}, V_x, V_y), E)$ is displayed in the graph view (\autoref{fig:screenshot}A). Here, each node has a Japanese label (omitted in all figures; English labels were superimposed as necessary). The primary attributes $V_x$ are displayed with the specified initial layout, i.e., \textit{star}, \textit{circular}, or \textit{linear} (\autoref{fig:layout}). The \textit{linear} layout aligns nodes horizontally. The \textit{circular} layout aligns nodes on a circumference. The \textit{star} layout places the node with the maximum degree at the center, and the remaining nodes surround the central node on a circumference. 
Then, $A$ and $V_y$ are visualized using a force-directed layout with ForceAtlas2 algorithm \citep{jacomy_forceatlas2_2014}. $V_x$ and $V_y$ are large nodes with different appearances. $A$ is visualized as small green nodes. Users wish to see what attributes that an applicant has, especially if the applicant looks outlier due to the location of nodes. A small pop-up appears to show the entire list of an applicant's attributes when users click $A$. Edge $E$ connects between $A$ and $V$. Due to the limitation of performance and perception, we do not recommend visualizing more than a hundred nodes at the same time. \textit{Limit of primary attributes} and \textit{offset of primary attibutes} parameters are useful for reducing the number of entire nodes. Since primary attributes are sorted by the occurrence of each attribute, users can retrieve subgraphs including an arbitrary range of primary attributes.

The force-directed layout calculates the positions of nodes based on physical simulations. Gravity makes two nodes closer, and repulsion makes two nodes more distant. All nodes are separated due to repulsion between nodes, but node pairs connected by an edge make closer. As more and more edges connect nodes, the connected nodes can be closer. As a result, the distance between attribute nodes relates to the number of edges, which helps users to see the relationship between attributes.

The primary attributes are anchored on the initial position while the secondary attributes and applicant nodes can move. Only users can move primary attribute nodes to an arbitrary position, which enables the users to perform manual clustering of nodes based on a category, feature, etc. When users move the position of primary attribute nodes, the position of the remaining nodes is recalculated and moves at the same time. Our initial implementation was to visualize all nodes using the force-directed layout without constraints on positions. However, the visualization between the primary or secondary attributes is obscure because the two attributes were superimposed or mixed. To avoid this, we suspend the position of the primary attribute. Further, the graph can be zoomed in/out with the mouse scroll. With the force-directed layout and zooming in/out, users can intuitively see correlations among the primary and secondary attributes.

The animated force-directed layout allows the users to observe the transition across different years. When users select a year to be displayed, applicant nodes and all edges are removed and re-rendered. We employ two techniques to preserve the mental map by reducing differences between timepoints. First, the positions of the primary and secondary attribute nodes are maintained at this time. Second, the applicant nodes are maintained if and only if the target node has the same edges over the years. Then, all edges are drawn between nodes, and the positions of secondary attributes and applicant nodes begin to change. This animation illustrates the dynamic transition across different years, which helps the users to see the trends. The transition is not limited to consecutive years, i.e., the users can observe dynamic differences between distant years. 

\paragraph{Configuration View} 

The configuration view (\autoref{fig:screenshot}B) allows users to select two types of attributes $V_x$ and $V_y$ from a dropdown menu. We assume that the users are HR specialists; thus, they select attributes based on their experience and insight. Here, the users select primary attributes $V_x$ and secondary attributes $V_y$. After that, $G(year, x, y) = ((A_{year}, V_x, V_y), E)$ is displayed in the graph view. The users can also specify other parameters to change the appearance of the graph view. We enumerate some of the parameters as follows:

\begin{itemize}
    \item \textit{initial layout}: Users can select the initial layout from \textit{star}, \textit{circular}, or \textit{linear} (\autoref{fig:layout})
    \item \textit{year}: Users must specify the fiscal year (FY) to visualize using a slider or input text.
    \item \textit{limit of primary attributes}: Users can limit the number of primary attributes nodes to display.
    \item \textit{offset of primary attributes}: Users can fetch the primary attributes nodes skipping the specified number of attributes.
    \item \textit{auto play}: When users enable autoplay mode, the dynamic transition of the graph is automatically played.
\end{itemize}

\paragraph{Chart View}

After the users specify two types of attributes, they can observe the time series of only the primary attributes. Here, line charts for each year are shown. The line chart in \autoref{fig:screenshot}C shows the transition of the number of applicants with the primary attribute. In this case, X-axis is the time series (FY 2014-2020), and the Y-axis is the node degree of the primary attribute. The chart view allows users to focus on the transition of years that is often prominent in the line chart.

\section{Case Studies}

We now apply the proposed system on the real-world dataset, an HR applicant data. Applicant data in the FY 2014-2020 were provided by \proposedcompany Corporation. The data were anonymized, and all personal information was removed or masked. Further, all analyses were performed on a \proposedcompany Corporation's internal workstation. The applicant data were dumped in isolated comma-separated values files for each fiscal year. The total number of users who registered with the system was about $400,000$; however, most registered to the company's system but did not apply for a position in the company. The total number of applicants that obtained employment was about $4,000$, and we focused on these data. The number of columns ranged from about $3,000$ to $7,000$ depending on the year, and only $1,100$ columns were shared across the seven-year period. That means other columns were not the same across the same period due to a change in a database schema. Therefore, we wrote custom scripts for matching columns that their names were changed across years.

Although the proposed system can support all enumerable values in a column, we selected 12 columns during user evaluation based on suggestions from HR specialists. HR specialists focused on the following six columns (applicant's attributes) in the case studies.

\begin{itemize}
    \item \textit{Applicant ID}: Universal unique id for all applicants.
    \item \textit{Location of university}: Nine regions (one oversea region and eight regions in Japan).
    \item \textit{Academic credential}: Bachelor, master, or doctoral student. Further, we also categorized new or previous graduates.
    \item \textit{Self-declared English skill}: Applicants select from [Entry, Conversational, Business, Native]-level English skills.
    \item \textit{Job field}: Jobs at \proposedcompany Corporation (sales, system engineer, research, HR, etc.) for which applicants applied.
    \item \textit{Internship history}: Company names where applicants worked as interns.
\end{itemize}

\begin{figure}[t!]
 \centering 
 \includegraphics[height=0.9\textheight]{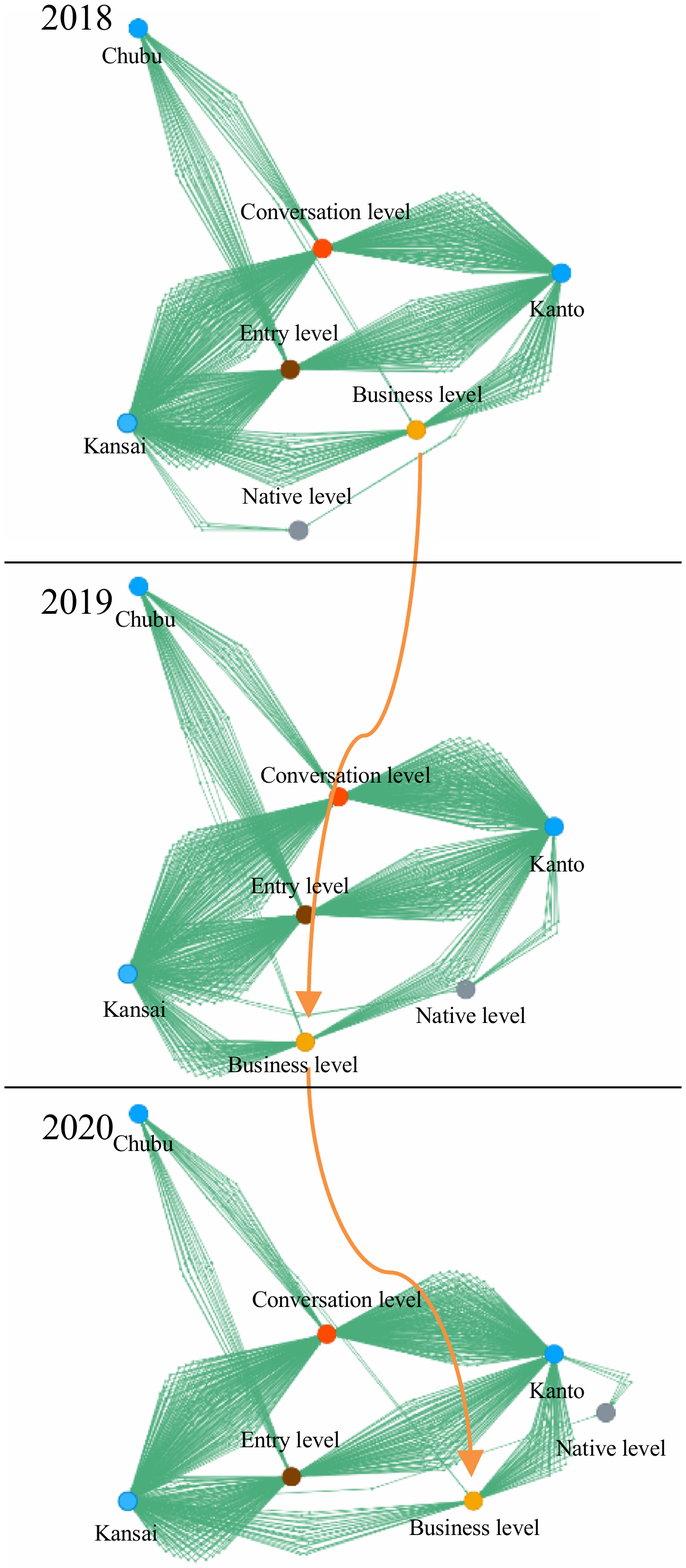}
 \caption{Example showing \textit{location of university} versus \textit{self-declared English skill} from FY 2018 to 2020.}
 \label{fig:example1}
\end{figure}

\subsection{Example A. Location of university versus self-declared English skill}

We present the first case study to demonstrate the transition over three years using the animated force-layout rendering in \proposedmethodz. \autoref{fig:example1} shows the graph view between the \textit{location of university} versus \textit{self-declared English skill} of prospective employees. Here, the three large blue nodes correspond to three regions in Japan, i.e., Kanto, Kansai, and Chubu\footnote{Kanto is the central-eastern part of Japan's main island and includes Tokyo. Kansai is the central-western area and includes Osaka. Chubu is the central area and includes Nagoya.}. We selected the \textit{circular} layout for this case. The large colored nodes (except for the blue nodes) are self-declared English skills collected from the submitted resumes, and the small green nodes are prospective employees. In this case, the colors differ based on the applicant's level of English: brown is entry-level, orange is conversational, yellow is business, and gray is native. In FY 2018 and 2020, the yellow node (business level English skills) is generally located between two blue nodes. However, the yellow node moved from the blue node Kanto to the blue node Kansai in FY 2019. This shows the trend in FY 2019 that there were more employees with business-level English skills in Kansai, but fewer applicants had the same skill in Kanto and Chubu region. Further, the gray node moves dynamically because its degree is less than that of the other large nodes. We also observe an attribute with a small number of applicants, which is often ignored by quantitative analyses. This result is an example to observe trends via the time-varying graph visualization.

\subsection{Example B. Major in university (liberal arts or sciences) versus job field}

\begin{figure}[t!]
 \centering % avoid the use of \begin{center}...\end{center} and use \centering instead (more compact)
 \includegraphics[height=0.9\textheight]{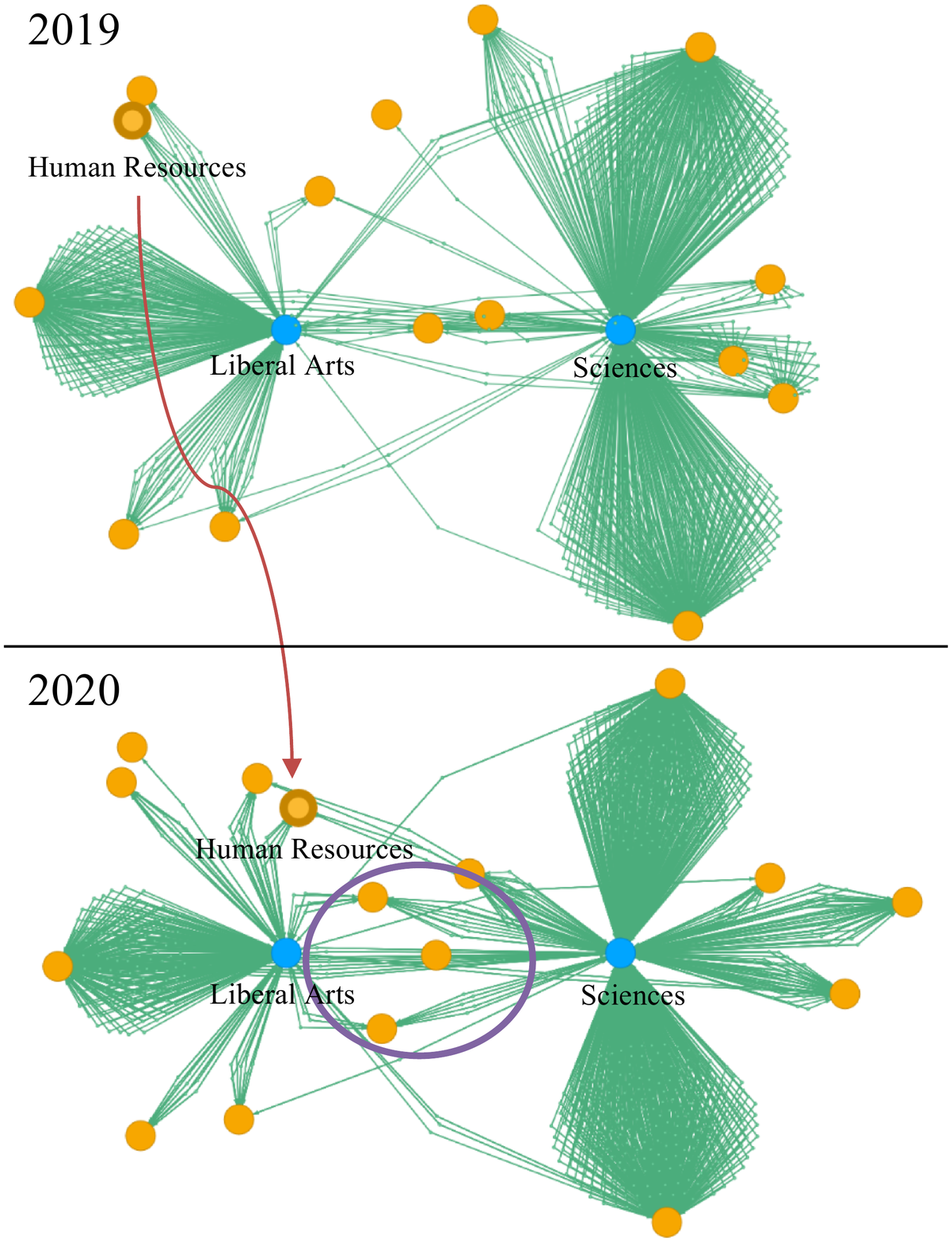}
 \caption{Example showing the transition of \textit{major in university} versus \textit{job field} from FY 2019 to 2020.}
 \label{fig:example3}
\end{figure}

\autoref{fig:example3} shows an example of integration with prior knowledge of HR specialists and the corresponding visualization. 
Traditionally, the selection process in most companies in Japan depends on the student's future occupational class. Students who major in social sciences, law, or humanities (referred to as liberal arts) will enter administrative jobs such as accounting, HR, personnel, sales, or purchasing. Students majoring in engineering or sciences (referred to as sciences) will enter technical jobs \citep{pucik_white-collar_1984}. They are assigned the department after job training in a company for several months. \proposedcompany Corporation had employed applicants either administrative or technical jobs, mainly. Recently, \proposedcompany Corporation has started another recruitment policy where each department directly hires applicants.
This example shows the transition of \textit{major in university} versus \textit{job field} from FY 2019 to 2020.

Here, the two blue nodes are \textit{major in university} (liberal arts or sciences). The highlighted orange node is the HR department. The blue node connected to the highlighted orange node in FY 2019 is the liberal arts node. This indicates that all new employees who assigned to the HR department in FY 2019 majored in liberal arts. However, in FY 2020, the highlighted orange node moves to the middle of the two blue nodes (shown by the red arrow), which indicates that the HR department began to employ people who majored in sciences or engineering. Similarly, three nodes are shown between the two blue nodes (shown in the purple circle). These three job fields employed people who majored in either liberal arts or sciences. The number of orange nodes located between the two blue nodes indicates that an increasing number of job fields tends to employ applicants regardless of their university major. Note that this trend does not mean that university majors are not considered. Instead, this represents the fact that diverse expertise becomes in demand in various job fields. These results show an example to observe the change in recruitment policies for job fields.

\subsection{Example C. Academic credential versus self-declared English skill}

\begin{figure}[t!]
 \centering % avoid the use of \begin{center}...\end{center} and use \centering instead (more compact)
 \includegraphics[height=0.85\textheight]{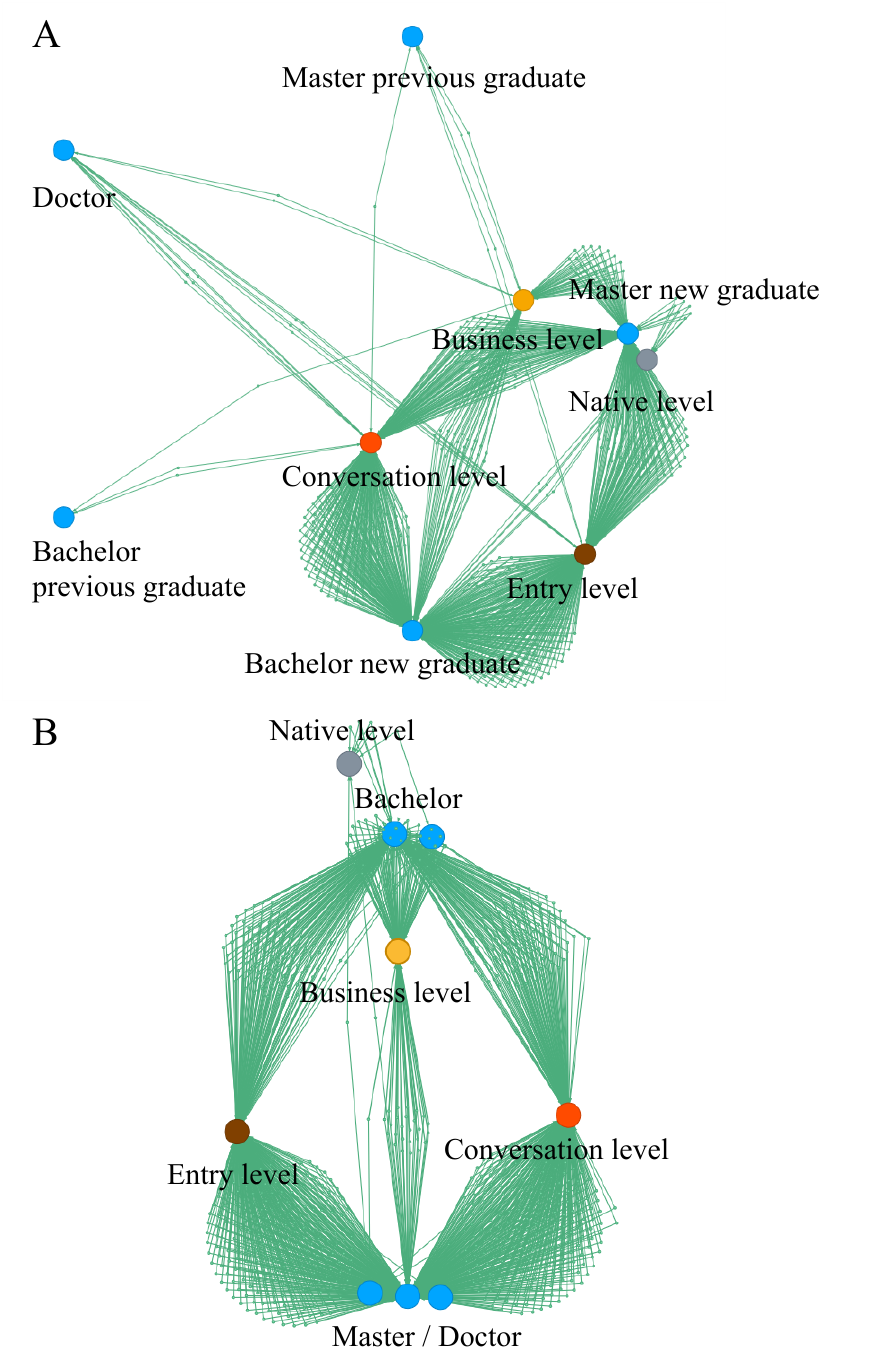}
 \caption{Example showing \textit{academic credential} versus \textit{self-declared English skill} of prospective employees: (A) Initial \textit{star} layout and (B) layout after manual clustering.}
 \label{fig:example2}
\end{figure}

\autoref{fig:example2} shows an example of the manual clustering of primary attributes. Here, the primary attributes are \textit{academic credential}, and the secondary attributes are \textit{self-declared English skills}. In the original data, the \textit{academic credential} fields store five values, namely, bachelor's degree new graduate, bachelor's previous graduate, master's new graduate, master's previous graduate, and doctoral graduate. This is because Japanese recruitment custom distinguishes applicants by degrees and by when they graduate. \autoref{fig:example2}A shows the initial layout that has five primary attributes (shown in blue) and four secondary attributes (shown in red, orange, gray, and brown). Due to the difference in the number of applicants in each primary attribute, relationships become difficult to interpret, which is a non-negligible burden for visual understanding.

\proposedmethod allows the users to easily move the positions of the primary nodes. HR specialists move the position of the primary nodes to two distant locations in \autoref{fig:example2}B. This procedure is a kind of clustering because the two distant positions can be seen as two clusters, i.e., bachelor and master/doctoral, which emphasizes the difference between degrees by ignoring the difference between new or previous graduates. Here, the top cluster is bachelor's, and the bottom cluster is master's/doctoral. The gray node is a native speaker, and the yellow node is a business-level English speaker. As can be seen, most native and business-level speakers have the academic credentials of bachelor. The trend is not unexpected because most masters/doctoral students are employed in technical jobs and most work in factories or research centers in Japan. In contrast, a large part of bachelor students are employed in administrative jobs, including international sales, purchasing, and HR. English skills are one of the criteria for assigning prospective employees to these positions. This example with the manual clustering function is more attractive for HR specialists. Users can also cluster graphs based on other criteria, e.g., whether they have already graduated. The proposed system can combine such empirical knowledge with data visualization.

\subsection{Expert Feedback}\label{sec:user}

We invited three HR specialists (denoted T1, T2, and T3) from various positions in the HR department as testers to evaluate the proposed system. To avoid bias, we invited several HR specialists, different from those who participated in the previously discussed biweekly meetings. The user evaluation phase lasted an hour for each person. 
We made a presentation (this lasted for 20 minutes) explaining how to use the proposed system, and then we watched how the users tested this system and interviewed the user (this lasted for 40 minutes). We observed that the users operated the system without special training. The questions we asked the users during the interview are as follows.

\begin{enumerate}
\renewcommand{\theenumi}{(Q\arabic{enumi})}
    \setlength{\leftskip}{0.2cm}
    \item How did you feel when you used the system? (General usability)
    \item Which part of the system was helpful for information discovery? (Visual design)
    \item Which part of the system did you find unfamiliar to use at first instance? (System usability)
    \item What additional features do you think could be integrated into the system to help with further knowledge discovery? (Additional data source)
    \item What did you discover using this system? (Knowledge discovery)
\end{enumerate}

We summarized the testers' responses as follows:

\begin{enumerate}
\renewcommand{\theenumi}{(A\arabic{enumi})}
    \setlength{\leftskip}{0.2cm}
    \item They had a good impression of the system because this kind of visualization is novel to them, and they could intuitively understand the relationship between attributes.
    \item The graph view is the most important component because the topology and transition of the graph are displayed simultaneously, which helps them recognize trends. In contrast, they did not find the chart view interesting.
    \item What surprised the testers the most was that the primary nodes can move everywhere, although several testers did not recognize this function at first. With this function, the testers clustered primary attributes based on their criteria. For example, a university can be categorized by grade or location depending on what the tester wishes to observe. Thus, clustering nodes based on individual criteria attracted more interest than the provided layout.
    \item After the user evaluation, the testers requested to import other data. For example, they would like to visualize the time-varying visualization on the snapshot of each day, which means the transition of the daily trend during a single recruitment period. The daily log must be stored in the database to track daily change. However, such data are currently not stored in the database because it only stores finalized information. As a result, the testers realized that they must have a log tracking system during the recruitment period.
    \item They all achieved knowledge discovery through the system, as described in several case studies in this section. 
\end{enumerate}

The testers noticed two issues with the attribute normalization. (1) Several job fields shared similar names disappeared and appeared at the same time during the transition between two years. However, some of these appear to be artifacts. Indeed, the restructuring of the organization has occasionally led to a change in the names of several divisions. (2) Unnormalized data were stored in the database. Even in a snapshot of a specified year, they observed that there were several different words representing the same object in the internship history attribute.
This issue is often referred to as ``employer normalization'' \citep{liu_companydepot:_2016, liu_supporting_2017}, which requires effort to annotate the different name entities to be the same. This unexpected appearance of the graph also suggests there is room for improvement to highlight the disappearing or appearing nodes like TempoVis or GraphDiaries \citep{ahn_temporal_2011,bach_graphdiaries:_2014}. Through the proposed system, unnormalized data could be detected via an unexpected appearance in the graph layout.

We summarized the comments from each tester as follows. T1, who is a section manager in the recruitment branding team, appreciated the overview of two specified attributes, which mitigated their burden compared to the conventional use of spreadsheets or BI tools. T1 suggested that the system would be useful for continuous improvement of employment policies because the system could visualize trends from all available data.
T2, who is a chief of the recruitment team for technical jobs, highlighted that the number of the edges corresponds to the correlation between two attributes, which provides clues on how to inspect these two attributes extensively. 
T3, who was just assigned to the HR department three months ago, easily caught up with the latest recruitment trends. T3 highlighted that the dynamic visualization of time-varying graphs shows the volatility of each attribute, which contributes to recognizing changes in trends.
These comments indicate that the system is widely accepted by the testers regardless of their expertise or position.

\section{Discussion}

\proposedmethod system proposes a time-varying multi-partite graph model for converting tabular data to the graph to apply dynamic graph visualization methods. Existing data-wrangling methods were not sufficient to generalize a graph model and visualization to temporal heterogeneous tabular data. Therefore, we modeled the multi-partite time-varying graph model and adapted the dynamic graph visualization to that graph. This is a remarkable user study for the integrated system of the time-varying graph model converted from heterogeneous HR data and visualization designed for that model. This result also underscores the pressing need for a general data wrangling framework of the time-varying graph model and dynamic visualization for temporal tabular data.

One of the main goals of \proposedmethod system is to provide HR specialists with an interactive visualization so that they can discover the time-varying relationship on applicants' history. The multi-partite graph visualization enabled HR specialists to focus both on temporal changes of trends and attributes of each applicant.
The combination of visualization techniques is novel for HR data analysis and enables HR specialists to gain new insights. Most importantly, HR specialists noticed that several columns are stored in undesirable ways. Such observations would contribute to improving the recruitment management system for next year's recruitment process. The system is a design guide for the future development of an HR data analysis system.

The categorical characteristics of the HR database and the schema-less graph database underscore the importance to handle HR data as a graph. In general, relational databases require a schema, whereas graph databases do not require a schema in advance. Thus, there is no need to migrate the graph database schema even if the table schema will be changed next year. The system can visualize new data as soon as the new data is imported into the system. We also expect that the system can be integrated with other relational databases.

We added several constraints to the visualization to make the system more interactive and intuitive, and these constraints were needed. The maximum number of applicant nodes shown at the same time would not be exceeded to a few hundred. Since applicants vary each year, applicants from different years are not displayed at the same time. Therefore, the system can visualize at most a few thousands of applicants in total duration. However, prospective employees of new graduates are a few thousand in total on \proposedcompany Corporation. For the initial analysis, the system is found to be sufficient. Further, we limited the number of attributes that can be observed simultaneously to two. Several relationships can exist between three or more attributes; however, these relationships are reduced to a one-to-one relationship between two attributes. As a result, finding the characteristics between attributes is possible.

\subsection{Limitations and Future Works}

We demonstrated \proposedmethodz's usability in a large Japanese company. The system can also be extended to other countries and other large companies with a closer number of prospective employees. However, there is still room for improvement for all applicants to be inspected. Note that applicants are selected and filtered out as the recruitment process proceeds. As a result, the database becomes sparse, which suits for the graph data structure more than tables. Meanwhile, we need to explore the visualization technique for a larger number of applicants. Furthermore, we need to explore better shape and color of attribute nodes if the number of attributes increases. Nevertheless, the scalability of the pre-processing and backend system is maintained even for future visualization updates owing to graph databases and modeling.

To employ other visualization techniques on top of \proposedmethod system is possible. Applicant nodes are useful for understanding the number of applicants, but this is sometimes redundant and raises performance issues. One way to mitigate this is by removing applicant nodes and connecting attribute nodes directly. Since the proposed system supports multi-column attributes, removing applicant nodes makes visualization difficult. If the degree of applicant nodes is restricted to two, we can easily remove the node and connect adjacent nodes using a single edge. However, removing a node whose degree is greater than three generates a hypergraph \citep{bretto_hypergraph_2013} whose edge connects greater than two nodes. In fact, if there is an inflated array such as \textit{internship history} with greater than two items, the graph becomes a hypergraph by definition. Although a hypergraph is the generalization of the graph used here, rendering a hypergraph is more complicated than the current method. Thus, we do not employ a hypergraph approach. The other way is bundling the same edges by merging applicant nodes whose edges have the same destination, like Parallel Sets \citep{kosara_parallel_2006} or RadialNets \citep{alsallakh_radial_2013}. However, both methods lose a simple interface to see each applicant's information when users click each node. There is room for exploring more effective visualization.

We showed that HR specialists can find trends from 12 columns. The data model and pre-processing pipeline can convert the entire columns into a multi-partite graph. However, the efficient way of selecting two attributes from more than thousands of columns has not been implemented, and the efficient way to provide this is not known. For example, categorizing names of columns requires the HR expert's prior knowledge or natural language processing. At the moment, we display only 12 columns by depending on HR specialists' prudent choice in advance. We tried to add a visualization module for selecting the attributes, but however, we found that it would not be effective without any categorization by HR specialists in advance. We plan to explore an effective recommendation method to help HR specialists select columns.

We succeeded to observe a few years' trends in the case studies. Since we aimed to visualize a long-time range transition, we applied the animation method to visualize the entire duration. Unfortunately, HR specialists were unable to observe any drastic transition in the trends throughout the entire duration using the system. We suspect that there was hardly any drastic transition of employment trends in \proposedcompany Corporation probably because the economic growth in Japan was stable and thus the unemployment rate was kept low \citep{stat62:online}.
Further, we did not evaluate the recruitment policies during the research period because the recruitment period in Japan is limited to once per year. These concepts are reserved for future work in the field of HR management.

One potential application of interactive graph clustering would be in tackling the problem of name identification such as employer normalization. Previous studies have employed heuristic and machine learning methods with manual annotation, web resource, and business database \citep{liu_companydepot:_2016, liu_supporting_2017}. Still, continuous manual effort is required. Name variants drop the reliability of analysis; however, curation for normalizing such entities is often time-consuming. Since unexpected data can be found more easily with the visualization, we suggest that further research should be undertaken in an interactive name identification system using graph visualization.

\section{Conclusion}

This study has proposed \proposedmethod system for HR specialists, which visualizes time-varying graphs from a heterogeneous table for the simultaneous recruitment of new graduates. To the best of our knowledge, this study is the first attempt to display large heterogeneous HR tabular data as a dynamic graph. Three case studies demonstrate the usability of time-varying graph animation with an interactive interface for inspecting heterogeneous databases. Future work should involve a general framework with a time-varying-graph-based data model and dynamic visualization for heterogeneous temporal tabular data.


\section*{Declaration of competing interest}

The authors declare that they have no conflict of interest.

\section*{Acknowledgements}
The authors thank Hideki Sugiyama, Kentaro Kuroda, and Kimio Minami from \proposedcompany Corporation for their support on requirement collection.

\bibliographystyle{elsarticle-harv} 
\bibliography{export-data}


\end{document}